\documentclass{article}

\PassOptionsToPackage{numbers, compress}{natbib}

\usepackage[preprint]{neurips_2023}

\usepackage[utf8]{inputenc} % allow utf-8 input
\usepackage[T1]{fontenc}    % use 8-bit T1 fonts
\usepackage{hyperref}       % hyperlinks
\usepackage{url}            % simple URL typesetting
\usepackage{booktabs}       % professional-quality tables
\usepackage{amsfonts}       % blackboard math symbols
\usepackage{amsmath}
\usepackage{graphicx}
\usepackage{nicefrac}       % compact symbols for 1/2, etc.
\usepackage{mathtools}
\usepackage{microtype}      % microtypography
\usepackage{xcolor}         % colors
\usepackage{comment}
\usepackage{wrapfig}

\usepackage{algorithm}
\usepackage{algpseudocode}
\usepackage{mathrsfs}
\graphicspath{ {images/} }

\usepackage{mathtools}
\DeclarePairedDelimiterX{\infdivx}[2]{(}{)}{%
  #1\;\delimsize\|\;#2%
}

\usepackage{amsthm}
\newtheorem{lem}{Lemma}

\newtheorem{defn}{Definition}

\newtheorem{prop}{Proposition}
\title{A General Space of Belief Updates for Model Misspecification in Bayesian Networks}

\author{
  Tianjin Li%\thanks{Use footnote for providing further informationabout author (webpage, alternative address)---\emph{not} for acknowledgingfunding agencies.} 
  \\
  Nuffield Department of Clinical Neurosciences\\
  University of Oxford, UK\\
  \texttt{tianjin.li@hertford.ox.ac.uk} \\
}

\usepackage{comment}
\newcommand*{\doubleunderline}[1]{\underline{\underline{#1}}}
\begin{document}
\maketitle
\begin{abstract}
 
%\section{Motivation}
In an \textit{ideal} setting for Bayesian agents, a perfect description of the rules of the environment (i.e., the objective observation model) is available, allowing them to reason through the Bayesian posterior to update their beliefs in an optimal way. But such an \textit{ideal} setting hardly ever exists in the natural world, so agents have to make do with reasoning about \textit{how they should update their beliefs} simultaneously. This introduces a number of related challenges for a number of research areas: (1) For Bayesian statistics, this deviation of the subjective model from the true data-generating mechanism is termed \textit{model misspecification} in the literature. (2) For neuroscience, it introduces the necessity to model how the agents' belief updates (how they use evidence to update their belief) \textit{and} how their belief changes over time. The current paper addresses these two challenges by (a) providing a general class of posteriors/belief updates called \textit{cut-posteriors} of Bayesian networks that have a much greater expressivity, and (b) parameterizing the space of possible posteriors to make meta-learning (i.e., choosing the belief update from this space in a principled manner) possible. For (a), it is noteworthy that any cut-posterior has \textit{local}\footnote{local in the sense that all computations in computing the posterior are \textit{within that particular module}} computation only, making computation tractable for human or artificial agents. For (b), a Markov Chain Monte Carlo algorithm to perform such meta-learning will be sketched here, though it is only an illustration and but no means the only possible meta-learning procedure possible for the space of cut-posteriors. Operationally, this work gives a general algorithm to take in \textit{an arbitrary Bayesian network} and output \textit{all possible cut-posteriors in the space}.
   
\end{abstract}

\section{Introduction}
\subsection*{Motivations}
The use of a probabilistic graphical model (or PGM) $G$, obtained from taking the union of a set of smaller subgraphs $(M_i)_{i\geq 1}$ and identifying overlapping vertices as the same, is becoming increasingly popular as more computational power is available and interdisciplinary research flourishes. In reality, each of these subgraphs, or \textit{module}, may come from a different discipline of study, or be built using the expertise of a specific data modality. 
\begin{comment}
    As an example of this approach, a model that relates human health to air pollution can be built from a module that predicts air pollution with climate data \textit{and} a module for human health based on medical sciences and health records (Blangiardo et al., 2011).\par
\end{comment}
As \cite{jacob2017better} points out, the conventional Bayesian posterior that updates the entirety of $G$ `simultaneously' is shown to be optimal when $G$ perfectly describes the underlying data-generating mechanisms \cite{bernardo2009bayesian}. The issue is that it is hardly possible to build models that perfectly describe reality using even the best techniques available. The almost inevitable deviation of our subjective model from the true data-generating mechanism is termed \textit{model misspecification} in the literature \cite{nott2023bayesian}. And, when conventional Bayesian posterior is used, even misspecification in a small part of the model (in a single module) is known to spread and cause misspecification in the full model so long as information flow in and out between any pair of modules \cite{liu2009,Grünwald2012}.\par
To address this concern, \textit{cut-posteriors} (cut models) are introduced, with the idea being to \textit{cut} the information flow from the suspected modules to the modules we trust, or operationally, to update the trusted modules before the suspected modules \cite{plummer2015, Bissiri2016,jacob2017better}. But, as an active area of development, cut-posteriors so far developed are limited in scope and formality. In some, part of data is thrown away altogether due to the cut, and in others, users of the model are forced to rely on heuristics to build even one cut posterior, making model selection for cut models underpowered.\par
In light of these ongoing issues in current cut model development, we propose a way to formally define: (a) the general space of possible cut-posteriors in any Bayesian network, (b) a way to efficiently parameterize that space, and (c) an algorithm to take in \textit{an arbitrary Bayesian network} and output \textit{all possible cut-posteriors in the space}. The result is a flexible, computationally efficient framework that makes choosing a belief update given certain modeling assumptions easy. Moreover, the formulation of the cut-posterior space paves the way for automatic model averaging/selection\footnote{As an example, one MCMC algorithm is provided in Appendix \ref{appendix_randomwalk} that is computationally tractable for reasonably small Bayesian networks.}.\par

\subsection*{Layout}
Structurally, we shall start by giving the concepts and lemmas needed in section \ref{section_definitions}, before turning to the enumeration algorithms and the general space that form the gist of the paper in section \ref{enumeration_algorithms}. To put the theoretical framework to work, in section \ref{section_parameterization_example}, we shall discuss a step-by-step parameterization of the space of cut-posteriors with an extended example. where our proposed methods yield a better-suited posterior than existing cut models. Section \ref{section_parameterization_example} also demonstrates how to select good cut-posteriors depending on different nuanced modeling needs. Please note that sections \ref{section_definitions} and \ref{enumeration_algorithms} contain technical details that are \textit{not} needed to use the model. In order to use the framework of cut-posterior, section \ref{section_parameterization_example} is sufficient and can be read on its own. All supporting details that make this paper as self-contained as possible will be included in the appendices to make the main sections more focused.\par

\subsection*{Relation with Previous Works}
In this paper, the definition of \textit{modules} follows that of \cite{liu2022} as their definition allows local, within-module updates in a multi-module setting. In particular, \cite{liu2022} provides a \textit{general} framework for cut models by starting with two modules and then splitting them sequentially. While the end goal is similar here, the current work takes a more bottom-up approach to explicitly enumerate possible cut-posteriors via parameterization in a larger space, with some main differences being: (1) the construction provided here allows more flexibility---e.g., in what partitions into modules or conditional structures between modules are possible---and thus defines an even larger space of cut-posteriors; (2) an explicit parameterization of possible cut-posterior is given in the current work, making both enumerating the space and building a particular belief update simpler; (3) in terms of focus, unlike \cite{liu2022}, the current work focuses on 'cuts' that are between modules without delving into within-module \textit{cuts}. Thus, while many previous works such as \cite{plummer2015, Bissiri2016,jacob2017better} have discussed some alternative cut-posteriors, the generality and formality of our proposed space of belief updates are novel to our knowledge.\par

\section{Definitions---Modules, Module Graphs, and Decision Sets} \label{section_definitions}
\subsection{Modules and How to Merge Them}
Modules, as motivated in the introduction, typically come in the form of separate datasets, which form a subgraph of the overall graphical model, which we shall denote as $G$. Some nodes in $G$ function as collected, empirical data, while others are parameters that are updated. Throughout this paper, we shall assume $V(G)$ has a natural bipartition $(X, \Theta)$ and refer to `data nodes' and `(hidden) parameter nodes' without explicitly stating. \par 

For the purpose of forming cut-posteriors, it is paramount to be able to update parameters from data in their corresponding module(s) alone. For this reason, we define them as such that each module, considered as the subgraph of $G$, yields a well-defined posterior term for any subset of parameters in that module following \cite{liu2022}.
\begin{defn} \label{moduledef}
A set of modules $\mathcal{M}$ is defined with respect to a partition of $X$ (the set of all data nodes in $G$), $X^* = \{X_1^*,X_2^*,...\}$ which will form $\mathcal{M} = \{M_1,M_2,...\}$, respectively. To form module $M_i$, we start with $X_i^*$ and consider all directed paths that ends in some nodes $Y\in X_i^*$:
\begin{enumerate}
    \item if such a path contains no nodes from $X\setminus X_i^*$, we collect all of its vertices into $M_i$.
    \item if such a path does intersect $X\setminus X_i^*$ at somewhere, then the two must intersect at some $Z$ such that (sub-)path from $Z$ to $Y$ does not intersect any other vertices in $X\setminus X_i^*$. We collect all vertices in the (sub-)path, including $Z$, into $M_i$.
\end{enumerate}
\end{defn}

In particular, this definition means that the operation of merging modules in graph $G$ interacts well with merging underlying subsets in the partition, which we formalize as the following:

\begin{prop} \label{prop_unique_module}
For any $I \subseteq X^* = \{X_1^*,X_2^*,...\}$, let $\mathcal{M}_I \subseteq \mathcal{M}$ denote the corresponding set of modules that is formed from $I$ according to Definition \ref{moduledef}. Then, if $M$ is the module obtained from applying Definition \ref{moduledef} to $\bigcup I$, we have that 
\begin{equation}
    \bigcup \mathcal{M}_I = M
\end{equation}
\end{prop}

Prop. \ref{prop_unique_module} is what allows us to enumerate partitions of $X$ instead when enumerating possible sets of modules, $\mathcal{M}$. In addition, however, this means that we can merge modules when appropriate to reduce the number of modules, thus reducing the size of space of cut-posteriors that we will form eventually. 

\subsection{Undirected and Directed Module Graphs} \label{section_module_graphs}
Being able to update any parameter $\theta$ within any module $\theta$ is a member of allows us to perform a full update on $G$ by updating one module at a time, following a linear order. This order on modules is specified by module graphs.\par
After specifying a set of modules $\mathcal{M}$, we now define \textit{undirected} and \textit{directed module graphs} formally. Module graphs essentially are graphical representations of the modular structure, so in both cases, we represent each module as a vertex and use edges (undirected or directed) to capture the interaction between them.

\begin{defn} \label{modulegraphdef}
Given $G=(\Psi,E)$ and a set of modules $\mathcal{M}$, we obtain the corresponding undirected module graph $H_{(G,\mathcal{M})}$ by taking a vertex $v_M$ for each module $M\in \mathcal{M}$ and having an edge between $v_A$ and $v_B$ if and only if two modules $A$, $B\in \mathcal{M}$ intersect at some vertices in $G$. When $(G,\mathcal{M})$ is clear from the context, we abbreviate $H_{(G,\mathcal{M})}$ to $H$ for simplicity.
\end{defn}
Undirected module graphs capture the constraints that $G$ and $\mathcal{M}$ impose on a modular structure but do not specify what the exact structure is. To do that, we need directed module graphs.

\begin{defn}
Given an undirected module graph $H$, we call any acyclic orientation (i.e., any directed acyclic graph obtained by assigning a direction to each edge in an undirected graph) $G_\mathcal{M}$ of $H$ a directed module graph.
\end{defn}
This definition implies that we can have many directed module graphs for the same undirected module graph. And, for a particular directed module graph, each directed edge $(v_i, v_j) \in E(G_\mathcal{M})$ means that we update the corresponding module of $v_i$ before (though not necessarily immediately before) that of $v_j$. The full linear order that precisely describes the sequence of modules we update is then represented as a \textit{topological ordering} $S$ of $V(G_\mathcal{M})$ (or equivalently $\mathcal{M}$). As any directed module graph is a DAG by definition, there is at least one such $S$, meaning there is at least one viable linear order available.

\subsection{Decisions and Decision Sets} \label{decision_sets}

Roughly, a $decision$ refers to a number of updating choices we make regarding a parameter $\theta$, e.g., which module to update it in. The decisions for any $\theta$ that belongs to exactly one module (denoted as $\theta \in \doubleunderline{\Theta}$) are trivial, so we will ignore those. A $decision$ $set$ is an object containing all the $decisions$ that we as modelers make for each shared parameter (a parameter that is in more than one module, denoted as $\theta \in \underline{\Theta}$). This decision set completes the information needed to determine a specific cut-posterior.\par

For notations, given a graphical model $G$, a set of modules $\mathcal{M}$, a directed module graph $G_\mathcal{M}$, and a topological ordering $S$ of $\mathcal{M}$ on $G_\mathcal{M}$, we relabel the elements of $\mathcal{M}$ as $\{M^{(1)}, M^{(2)},..., M^{(\lvert\mathcal{M}\rvert)}\}$ according to $S$. For a given $\theta \in \Theta$, we denote the subset of modules that $\theta$ belongs to as $\mathcal{M}_\theta$. We also relabel the elements of $\mathcal{M}_\theta$ as $\{M_\theta^{(1)}, M_\theta^{(2)},..., M_\theta^{(\lvert\mathcal{M}_\theta\rvert)}\}$ according to $S$. From here on, we shall refer to specific modules by their new labels without specifying when there is a topological ordering $S$ clear from the context.

\begin{defn}
Given graph $G$, a directed module graph $G_\mathcal{M}$, and a topological ordering $S$ of $\mathcal{M}$, a decision for any parameter $\theta \in \Theta$ is an ordered pair $(D_\theta, x_\theta)$ with $D_\theta$ a bipartite graph with partition $\{T_\theta, C_\theta\}$ and $x_\theta$ an integer between $1$ and $\lvert\mathcal{M}_\theta\rvert$, that satisfies all of the following:
\begin{enumerate}
    \item $D_\theta$ has vertex set $V(D_\theta) = (v_1, v_2,..., v_{\lvert\mathcal{M}\rvert})$, each representing the corresponding module in $\mathcal{M}_\theta$ ranked according to $S$.
    \item $v_1$, $v_{x_\theta} \in T_\theta$.
    \item For any $v_i \in C_\theta$, it has precisely one neighbour, and its only neighbour $v_j$ must satisfy that $j < i$.
\end{enumerate}
Also, a collection of one such ordered pair for every $\theta \in \Theta$ is called a decision set, denoted by $\mathcal{D}$.
 
\end{defn}
This definition means that we choose to update $\theta$ in $x_\theta$-th module (ordered according to $S$) that it appears in. And, when choosing what to do with $\theta$ in the $i$-th module $M$, if $v_i \in T_\theta \setminus v_{x_\theta}$, we create a new tilde parameter for $\theta$ and update it instead. On the other hand, if $v_i \in C_\theta$, we condition on the version of $\theta$ created when updating the module represented by the unique neighbor of $v_i$.

\section{Outputting possible cut-posteriors for arbitrary Bayesian networks} \label{enumeration_algorithms}
We now give the algorithms that sequentially enumerate the possibilities at each stage. The first takes a graphical model $G$ and a partition $\mathcal{X}$ of $X$ as arguments and outputs all possible sets of modules $\mathcal{M}$. The second takes a particular set of modules $\mathcal{M}$ (and $G$) as arguments and outputs all possible directed module graphs $G_\mathcal{M}$; the third then takes a particular directed module graph $G_\mathcal{M}$ (and $G$) as arguments and outputs all possible decision sets $D$ with respect to a topological ordering $S$ on $\mathcal{M}$; the last then takes a directed module graph $G_\mathcal{M}$ and an ordered pair $(S,D)$ (and $G$) and outputs a particular posterior, written in explicit algebraic form. Note that the first two are standard problems in programming and thus have many well-known variants already. Thus, in this section we focus on the last two algorithms, which form the gist of the space of cut-posteriors.

\subsection{Algorithm 1: Enumerating Possible Decision Sets}
To enumerate possible decision sets for a directed module graph, we first perform Depth First Search on $G_\mathcal{M}$, starting with any vertex without a parent, to obtain a topological ordering $S$.
\begin{prop}
For a fixed $\theta$, the number of possible partitions, $\{T_\theta, C_\theta\}$, is $2^{\lvert\mathcal{M}_\theta\rvert-1}$ since every vertex except $v_1$ can be in either $T_\theta$ or $C_\theta$.
For a fixed partition $\{T_\theta, C_\theta\}$, the total number of possible decisions $(D_\theta, x_\theta)$ is given by
\begin{equation}
\lvert T_\theta \rvert \cdot \prod_{v_i \in C_\theta}\lvert \{v_j \in T_\theta|j < i\}\rvert
\end{equation}
\end{prop}

The important thing is that possible decisions for $\theta$ \textit{do not depend on the decision for any other parameters}. Therefore, we can easily enumerate possible decisions by enumerating $\theta \in \Theta$, possible partitions, possible values of $x_\theta$, and possible unique neighbors of each $v_j \in C_\theta$ in a depth-first manner.

\subsection{Algorithm 2: Outputting Posteriors in Explicit Forms}
Before introducing the last algorithm, we shall first introduce a well-known graph theory property as a lemma without proof.
\begin{lem}
A directed graph is a DAG if and only if its vertices can be arranged as a linear ordering that is consistent with all edge directions (i.e., has a topological ordering).
\end{lem}

\begin{defn}[Algorithm 2]
Given graph $G$, a directed module graph $G_\mathcal{M}$, a topological ordering $S$ of $\mathcal{M}$, and a decision set $\mathcal{D}$, the algorithm is defined as follows:
\begin{enumerate}
    \item Repeat step 2, 3, and 4 for each $M_i$ ($1 \leq i \leq \lvert\mathcal{M}\rvert$) in order.
    \item Partition the union of $X_{M_i}^*$ and parameters of the current module $M_i$ and, $\Theta_{M_i}\cup X_{M_i}^*$, into three subsets, $U_{M_i}$ ($U$ for `to be updated'), $C_{M_i}$ ($C$ for `to be conditioned'), and $T_{M_i}$ ($T$ for `to be introduced as tilde parameters') according to the following:
    \begin{enumerate}
        \item If $v$ is a data node, then $v \in C_{M_i}$.
        \item If $v \in \doubleunderline{\Theta}_{M_i}$, then $v \in U_{M_i}$.
        \item Let $y_{iv}$ be such that $M_i$ is the $y_{iv}$-th module that $v$ appears in (according to $S$), then
        \begin{enumerate}
            \item If $y_{iv} = x_v$, then $v \in U_{M_i}$.
            \item Else if $v_{y_{iv}} \in T_v$, then $v \in T_{M_i}$.
            \item Else if $v_{y_{iv}} \in C_v$, then $v \in C_{M_i}$
        \end{enumerate}
        In particular, $T_v$, $C_v$, and $x_v$ here all refer to the decision in $\mathcal{D}$, a part of the argument.
    \end{enumerate}
    \item Let $\Theta_{C_{M_i}}'$ be the set that collects, for each parameter $\theta \in C_{M_i}$, the tilde version of $\theta$ created during updating module $M_j$, such that $v_j$ is the only neighbour of $v_{y_{i\theta}}$ in $D_\theta$. 
    \item If $U_{M_i} \neq \emptyset$, build update term for $M_i$ as:
    \begin{multline}
        p(\Theta_{U_{M_i}}, \tilde{\Theta}_{T_{M_i}}|X_{C_{M_i}}, \Theta_{C_{M_i}}') \propto p(\Theta_{U_{M_i}},\tilde{\Theta}_{T_{M_i}}, X_{C_{M_i}}, \Theta_{C_{M_i}}')\\ \propto p(\Theta_{U_{M_i}}|pa(\Theta_{U_{M_i}}))p(\tilde{\Theta}_{T_{M_i}}|pa(\Theta_{T_{M_i}}))p(X_{C_{M_i}}|pa(X_{C_{M_i}}))p(\Theta_{C_{M_i}}'|pa(\Theta_{C_{M_i}})
    \end{multline}
    where $X_{C_{M_i}} = C_{M_i} \cap X$ and $\Theta_{C_{M_i}} = C_{M_i} \cap \Theta$. Otherwise, let the term be $1$. \par Also, $\tilde{\Theta}_{T_{M_i}}$ always `creates' a new copy of any $\theta \in T_{M_i}$. In other words, if $\tilde{\theta}^{(2)}$ is already created and updated in a prior module, we include $\tilde{\theta}^{(3)}$ in $\tilde{\Theta}_{T_{M_i}}$ instead.
    \item Update all parameters that do not belong to any module, $\Theta_S$ through the term:
    \begin{equation}
        p(\Theta_S|pa(\Theta_S)) = p(\Theta_S|\Theta\setminus\Theta_S, X)
    \end{equation}
    \item Define the overall tilde-cut posterior as the product of all update terms built in step 4 and 5, with all tilde parameters ($\tilde{\Theta}$) integrated out with marginal priors:
    \begin{equation}
        p(\Theta_S|pa(\Theta_S))\int \prod_{i=1}^{\lvert\mathcal{M}\rvert}p(\Theta_{U_{M_i}},\tilde{\Theta}_{T_{M_i}}|X_{C_{M_i}}, \Theta_{C_{M_i}}')d\tilde{\Theta}
    \end{equation}
\end{enumerate}
\end{defn}

Note that in step 2, we partition $\Theta_{M_i}\cup X_{M_i}^*$ instead of the set of all vertices of $M_i$, omitting $X_{M_i}\setminus X_{M_i}^*$ in the process. The reason is that our definition of a module produces meaningful and well-defined posterior update terms so long as all variables of that module appear in it. This is satisfied when we update module $M_i$, as the only variables that do not explicitly appear are $X_{M_i}\setminus X_{M_i}^*$, but by Definition 1, for any such $x \in X_{M_i}\setminus X_{M_i}$ to be in module $M_i$, it must be a parent of some $\psi \in \Theta_{M_i}\cup X_{M_i}^*$, which means it is being conditioned on in one of the terms.

Also, since this algorithm picks just one out of potentially many orderings consistent with the directed module graph given and operates with that throughout, we want the algorithm to be able to output $any$ posteriors that are the outputs of other orderings, too. This is the content of the following proposition:
\begin{prop}\label{any_ordering_works}
Let $f_S:\Delta \to \mathcal{P}$ (with $\Delta$ being the space of all possible decisions for $S$) denote the function that takes a decision set to a posterior according to ordering $S$ and Algorithm 4, then $f_S$ is surjective.
\end{prop}

Note that it is equivalent to say that given the same set of modules $\mathcal{M}$ and directed module graph $G_\mathcal{M}$, the space of all possible cut-posteriors $\mathcal{P}$, is identical with the space of all possible posteriors given any specific topological ordering $S$. This makes $S$ redundant and consequential only for notation. \par

Also, proposition \ref{any_ordering_works} puts us in good shape to have the following, similar result about the uniqueness of enumerated posteriors.

\begin{prop}\label{post_unique}
Assume that for any $M_i \in \mathcal{M}$, $M_i \cap \doubleunderline{\Theta} \neq \emptyset$. Given a particular $G_\mathcal{M}$, an ordering $S$, and let $f_S$ be defined as in proposition \ref{any_ordering_works}, then $f_S$ is injective. Note that we define two posteriors, $p, p' \in \mathcal{P}$ are equal if and only if for any $M_i \in \mathcal{M}$, the update terms in $p$ and $p'$ are identical. 
\end{prop}
The assumption is in place to exclude the degenerate cases, namely when some modules have no intrinsic parameters. The definition of equivalence for posteriors are such in order to allow two posteriors to be equal precisely when the $underlying$ action in $G$ for all the module is the same, `underlying' here stresses that we do not refer to any specific labeling. 
Thus, for instance, proposition \ref{any_ordering_works} implies that for any ordering $S$ and any decision set $\mathcal{D}$, there exist $S'$ and $\mathcal{D}'$ such that $f_S(\mathcal{D}) = f_{S'}(\mathcal{D}')$.

\section{Parameterization of the space of cut-posteriors with a toy example} \label{section_parameterization_example}

\begin{wrapfigure}[31]{R}{0.45\textwidth}
    \centering
        \includegraphics[width = 0.45\textwidth]{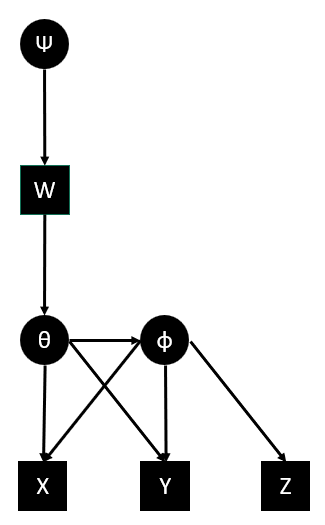}
    \caption{An example Bayesian network used throughout this section. Circular nodes refer to unobserved nodes/hidden parameters, while square nodes refer to data nodes.}
    \label{example_1}
\end{wrapfigure}
The enumeration algorithms that we have now established allow us to parameterize the space with three parameters, which is \textit{everything needed to specify a cut-posterior}.
To show how the results above are useful, this section illustrates how the space of cut-posteriors is parameterized \textit{and} how to choose them per the exact modeling needs in an example in fig. \ref{example_1}. It is important to note that these three parameters have a \textit{nested structure}---the value of the first parameter determines what values the second parameter can take, whose value in turn determines what values the third parameter can take. Note that here the three \textit{parameters} should be distinguished from the '(hidden) parameters' in the graphical model (the circular nodes in Fig. \ref{example_1}). For this reason, the latter kind should be referred to explicitly as 'hidden parameters' throughout this section.\par

\subsection{Parameter 1---Partition of the set of all data nodes}
\begin{figure}[t]
\begin{center}
\includegraphics[width = \textwidth]{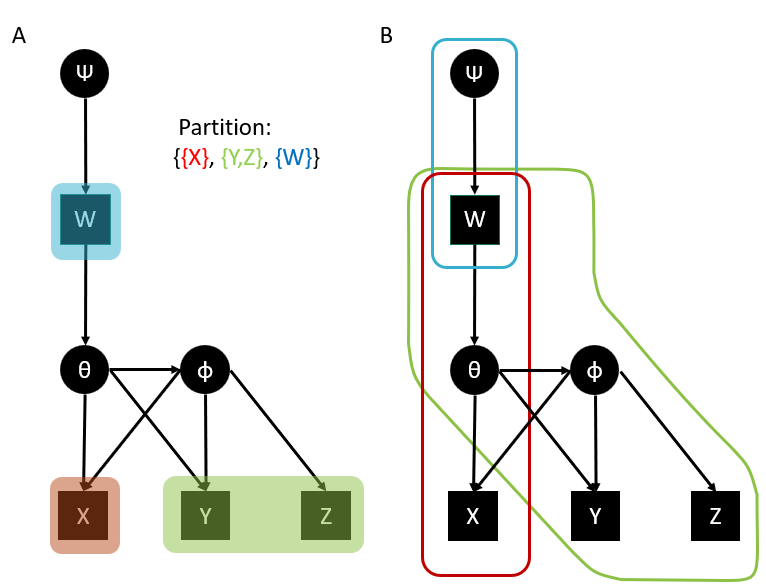}
\end{center}
\caption{Panel A: An example partition of data nodes. Panel B: Resulting division into modules according to module formation rules in Definition \ref{moduledef}}
\label{example_1.1}
\end{figure}

A \textit{partition} of a set is a grouping of its elements into non-empty subsets. The set of all data nodes in our running example is $\{X,Y,Z,W\}$, so an example partition is shown in fig. \ref{example_1.1}A. We can then apply definition \ref{moduledef} to determine what modules are in fig. \ref{example_1.1}B.\par

Crucial to note that for application purposes, this partition has the interpretation of \textit{defining modules as information sources to update unobserved nodes from}. This is because although there is some flexibility on which versions of hidden parameters to condition on (more on this later), cut-posterior dictates that we can only update one hidden parameter in \textit{one} particular module. Thus, by specifying the partition in fig. \ref{example_1.1}A, we may think that the data nodes $Y$ and $Z$ are qualitatively similar in some ways (e.g., reliability, data collection methods, etc.). For instance, if we are modeling multisensory integration of perception in cognitive science and $Y$ and $Z$ are both quantities relating to one sensory modality. \par

\subsection{Parameter 2---an linear order on modules}
\begin{figure}[!h]
\begin{center}
\includegraphics[width = 0.45\textwidth]{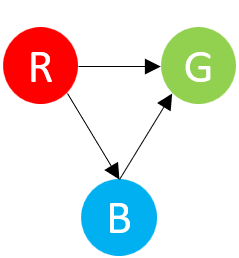}
\end{center}
\caption{An example of \textit{directed module graph} defined in section \ref{section_module_graphs}. Each node represents the module with the same color. The direction of edges specifies the precedence---$R \rightarrow G$ implies that the Red module should be updated \textit{before} the green module. Note that by prop. \ref{any_ordering_works} a directed module graph is sufficient to specify the second parameter since any linear ordering consistent with it induces an identical set of posteriors.}
\label{example_1.2}
\end{figure}

Operationally, cut-posteriors are formed by forming one multiplicative factor for each module in order and then multiplying them together. Thus, after determining what the modules are, the next step is to choose the linear order to update them in. One such order is shown in Fig. \ref{example_1.2}, where each node represents the module with corresponding color in \ref{example_1.1}B.\par

The linear update order on modules intuitively has the effect of preventing any misspecification from downstream modules to upstream modules. Here, we may choose the red module to be upstream may be due to that we are much more confident in the observation model for $X$, than its counterpart for $Y$, or $Z$, thus, we may prefer to update either $\theta$ or $\phi$ using $X$ \textit{only}.\par

\subsection{Parameter 3---decisions regarding individual hidden parameters}
After fixing on a linear order, the last parameter grants the user a lot of flexibility regarding individual hidden parameters. A hidden parameter can be updated in any of the modules that it is a member of, so decisions need to be made for each of the hidden parameters that appear in more than one module. For each hidden parameter's appearance in each module, we can either \textit{update} or \textit{condition} on it.\par
To illustrate these in the current example, $\theta$ appears in both the red \textit{and} the green module.
Having specified in fig. \ref{example_1.2} that the red module is more reliable than the green module, we may naturally want to update $\theta$ in the red module, this gives us the multiplicative factor \textcolor{red}{$p(\theta| W, X)$}. This still leaves what to do with $\theta$ in the green module open: we can choose to \textit{condition} on $\theta$ here, in order to leverage the information about $\theta$ learned in the red module to inform other \textit{updated} hidden parameters in the green module (in this case, $\phi$); this will produce a multiplicative factor of \textcolor{green}{$p(\phi|\theta, W,Y,Z)$}. On the other hand, we can also choose to \textit{update} $\theta$ along with $\phi$ in the green module, thus forming us a different factor \textcolor{green}{$p(\tilde{\theta},\phi|W,Y,Z)$} where we use $\tilde{\theta}$ to distinguish it from $\theta$, avoiding updating a hidden parameter twice.\par

Note that the factor for the blue module will always be \textcolor{blue}{$p(\psi|W)$} since it does not share any hidden parameter with any other module. Thus, depending on what our decision for $\theta$ in the green module is, we can assemble two cut-posteriors:
\begin{align}
    \pi_1(\theta,\phi,\psi|W,X,Y,Z) &= \textcolor{red}{p(\theta| W, X)}\textcolor{blue}{p(\psi|W)}\textcolor{green}{p(\phi|\theta, W,Y,Z)} \label{cut_pos_1}\\
    \pi_2(\theta,\phi,\psi|W,X,Y,Z) &= \int\textcolor{red}{p(\theta| W, X)}\textcolor{blue}{p(\psi|W)}\textcolor{green}{p(\tilde{\theta},\phi|W,Y,Z)}\pi(\tilde{\theta})d\tilde{\theta} \label{cut_pos_2}
\end{align}
Note that the marginal density $\pi(\tilde{\theta})$ is the same as marginal prior density $\pi(\theta)$. We integrate out the tilde variables in eq. \ref{cut_pos_2} because after all, only one version of $\theta$ should appear in the final cut-posterior reflecting our posterior belief about that variable. Thus, we need to \textit{decide} which version we want to keep (in this case, using $\theta$ likely more sense since the red module is more reliable).\par
\begin{figure}[t]
\begin{center}
\includegraphics[width = \textwidth]{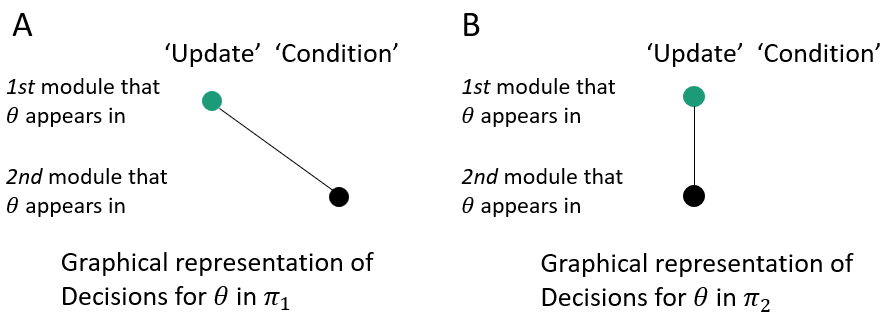}
\end{center}
\caption{An alternative equivalent representation of the decisions made in $\pi_1$ and $\pi_2$ using bipartite graphs. The two parts are the 'Update' and 'Condition' columns. Note that '1st' or '2nd' refers to the order of the module in the ordering specified in the second parameter. So, here, the '1st' and the '2nd' modules will be the red and the green modules, respectively. There will be exactly one cyan node in the \textit{`update' column}, designating the version of the hidden parameter that be kept in the overall posteriors. The technical definitions are specified in section \ref{decision_sets}}
\label{example_1.3}
\end{figure}

All these different decisions the framework offers in this subsection can seem difficult to keep track of, but in fact, all \textit{legal} sets of decisions have a unique representation in the form of a set of bipartite graphs. The technical details are specified in section \ref{decision_sets}, but for the two cut-posteriors we formed here, they respectively have a graphical representation shown in fig. \ref{example_1.3}. 
\section{Discussions}
\subsection*{Limitations \& future directions}
%On a very rough level, the fact that a cut-posterior updates a hidden parameter using some but not necessarily all sources of information (i.e., data nodes) closely mirrors the general idea behind attention. Both frameworks address the simple fact that \textit{the same observed quantity may not be relevant to an inferential goal in the same degree all the time}

The parameterization of the space of cut-posetiors paves the way to \textit{automate} the process of discovering and choosing one's belief update along with updating one's beliefs. This is made even simpler by needing to enumerate the key parameters instead. To illustrate this possibility, an example Monte Carlo Markov Chain (MCMC) algorithm is given in Appendix \ref{appendix_randomwalk}. The computation is tractable when the size of the cut-posterior space is small. \par

One way to realize belief update in Bayesian networks is as various message passing schemes \cite{parr2019}. Cut-posteriors are fundamentally inspired by the class of (often handcrafted) \textit{cut models} already in use as a partial solution to model misspecification \cite{plummer2015,jacob2017better}. Although not explored in the current work, it is plausible that the computations needed to realize cut-posteriors could be nicely captured by 'cutting' messages sent between modules. This connection can be key to an easy-to-implement algorithmic realization of cut-posteriors and at any rate, would be interesting for how cut-posteriors can be realized in neurobiological terms.\par

Another related issue in applying the cut-posterior framework to study how the brain infers is the definition of a module. The current work follows the definition of a module in \cite{liu2022} in order to make each module a functional \textit{sub-}Bayesian network in its own right, but such well-behaved modules are unlikely to coincide with the organization of the biological circuits. Although the lack of neuron-level implementation in the brain is a general problem for most, if not all, computational models of the brain, it is nevertheless interesting to see how the brain handles the problem of intractability, and whether it does so through a related sense of modularity as appears here.\par

It should be noted that the \textit{generality} of the cut-posterior framework is in comparison and thus by no means absolute: there are plenty of possible belief updates that do not belong to the space of cut-posteriors. One natural dimension to make the space even more general is the idea of 'incomplete cuts'---the idea that the information flow from one module to another could be neither all (Bayesian) nor nothing (cut). Such an idea is explored in what is termed \textit{semi-modular inference} (SMI) in \cite{2022carmona}. One benefit of introducing 'incomplete cuts' to the space of cut-posteriors is the sense of 'continuity' that could follow as a result: instead of needing to 'jump' in discrete steps in the current space, an agent may take a continuous walk in the abstract space of 'how to reason' through time.\par

\subsection{Conclusion}
Currently, the usefulness of Bayesian statistics (in particular, Bayesian networks) is hampered by the limitations in \textbf{flexibility} and \textbf{efficiency}. \par

Savage \cite{savage1972} famously differentiates between \textit{small-world} and \textit{large-world} settings for Bayesian inference---small-world is when all pertinent alternatives, outcomes, and probabilities are known while a large world problem is where either the prior, likelihood or both are uncertain \cite{binz2023metalearned}. For prior, a Bayesian agent can rely on the sheer amount of data to correct (given the right likelihood), but wrong likelihood, even in a small part of the Bayesian network, is known to spread and contaminate the inferential processes \cite{liu2009}. As small-world applies to very few problems. As hardly (if ever) are we situated within a small-world setting, it becomes a necessity to revise one's subjective model of the environment on the go. This is where limitation in \textbf{flexibility} becomes a challenge.\par

In light of this, note that one advantage of parameterizing the space of cut-posteriors is that it enables model selection or meta-learning. For instance, statisticians and neuroscientists can now construct a useful alternative posterior/belief update in a principled manner, by carefully choosing each parameter to suit one's nuanced modeling needs. But arguably what is even more impactful is the potential to \textit{automate} the process of discovering and choosing one's belief update along with updating one's beliefs, which is made possible now by enumerating the key parameters instead.\par

As for \textbf{efficiency}, one of the main challenges of Bayesian inference is its computational intractability. In fact, the number of steps needed to compute the normalization constant grows exponentially with the number of hidden parameters \cite{binz2023metalearned}. For comparison, the framework of cut-posteriors divides the graphical model into modules and performs computations inside individual modules only. This radically reduces the amount of computation needed so long as the number of hidden parameters and the number of modules is not trivially small. This tractability is particularly important for the brain, which needs to judiciously allocate its computational resources when updating a very large network of beliefs.\par

\section*{Acknowledgement}
Thanks to Geoff Nicholls, whose guidance and suggestions made this work possible in the first place; thanks to Yang Liu for sharing his dissertation through personal communications; and thanks to Junyu Ren for the helpful discussions and feedback.
\newpage
\bibliographystyle{plainnat}
\bibliography{bibliography}

\newpage
\appendix

\section{Automatic meta-learning: an example MCMC algorithm on the space of cut-posteriors} \label{appendix_randomwalk}
Now that we have formally defined the general posterior space and are equipped with algorithms to enumerate it, we demonstrate an important use of these concepts and tools by simulating a well-suited posterior with a random walk on the space. \par

In essence, we define a random walk on the posterior space, such that at each individual step, given $G_\mathcal{M}$, $S$, $\mathcal{D}$, and a resulting posterior $p$ we either
\begin{enumerate}
    \item perturb the directed module graph $G_\mathcal{M}$ by randomly merging two (adjacent) modules into one or splitting a module into two, then making accommodating changes to the decision set, or
    \item perturb a random vertex in $D_\theta$ for a randomly selected $\theta$.
\end{enumerate}
In doing so, we obtain a slightly different $G_\mathcal{M}'$, $S'$, and $\mathcal{D}'$, and these will jointly result in a slightly different posterior $p'$ in the posterior space, as shown in proposition \ref{post_unique}. The user can then run ELPD or other model selection methods to compare $p$ and $p'$; if the performance is better then we accept and proceed to the next step with $G_\mathcal{M}'$, $S'$, $\mathcal{D}'$, and $p'$ instead, otherwise we randomly accept or reject depending on how much worse the performance of $p'$ is. Since a plethora of such model comparison techniques exist, we shall leave the exact method to the user and focus instead on how to obtain a well-defined yet slightly different $p'$ at each step, as well as making sure the random walk is irreducible. \par

A few clarifications shall. Firstly, we identify all of the following and shall refer to them interchangeably: (a) a module in $\mathcal{M}$ (b) a vertex in $G_M$ (c) a set of vertices in $G$ obtained with module formation rules in Definition \ref{moduledef}. In addition, we sometimes abuse the notation to refer to a vertex, say $v_i$ in $D_{\theta}$, interchangeably with the module represented by that vertex. Notice here $v_i$ does not necessarily represent $M_i$, since although both are labeled with the topological ordering $S$, $V(D_\theta)$ is only a subset of $\mathcal{M}$ so indexing can be different. \par
Also, we identify different indexing that imply the same order of objects as the same ordering. In other words, it is equivalent to label modules $(A, B, C)$ as either $(M_0, M_{1.5}, M_4)$ or $(M_1, M_2, M_3)$, and we say one can always $adjusts$ the former to the latter canonical form.

\begin{defn} \label{randomwalk}
Given $G_\mathcal{M}$, $S$, and $\mathcal{D}$,

\begin{enumerate}
    \item with probability $q_0$, we perturb $\mathcal{D}$, or more specifically $(D_\theta, x_\theta)$, where $\theta$ is randomly drawn from $\underline{\Theta}$, and
    \begin{enumerate}
        \item with probability $q_1$, perturb $D_\theta$
        \begin{enumerate}
            \item with probability $q_2$, randomly sample any edge in $D_\theta$ that connects $v_i \in T_\theta$ and $v_j \in C_\theta$, then
            \begin{enumerate}
                \item with probability $q_3$, move $v_j$ to $T_\theta$ and delete the edge.
                \item with probability $1-q_3$, rewire the edge to connect $v_j$ and $v_k \in T_\theta$ instead ($v_k$ is randomly drawn from the vertices in $T_\theta$ with a smaller index than $j$).
            \end{enumerate}
            \item with probability $1-q_2$, sample any $v_i \in T_\theta$ such that $d(v_i)=0$, move it to $C_\theta$ and connect it to a randomly selected node in $T_\theta$ with a smaller index.
        \end{enumerate}
        \item with probability $1-q_1$, change $x_\theta$ to the index of a randomly selected vertex in $T_\theta$.
    \end{enumerate}
    \item with probability $1-q_0$, perturb $G_\mathcal{M}$, specifically
    \begin{enumerate}
        \item with probability $q_4$, sample any edge in $G_\mathcal{M}$, merge the two adjacent modules, say $M_i$ and $M_j$, by contracting their two vertices in the directed module graph. If this creates a cyclic graph, reject and sample again; otherwise, holding the indices of all other modules the same, randomly select an index out of all indices such that the resulting order can be \textit{adjusted} into a topological ordering of $G_\mathcal{M}'$, reject and sample again if there are no such indices. Call this ordering $S'$, modify the decision for any $\theta$ that belongs to both $M_i$ and $M_j$ according to the following:
        \begin{enumerate}
            \item If the corresponding vertices of both $M_i$ and $M_j$ belong to $T_\theta$, simply contract those two vertices in $D_\theta$.
            \item If the corresponding vertices of both $M_i$ and $M_j$ belong to $C_\theta$, contract those vertices, and if the resulting degree has two neighbors, randomly remove a neighbor by deleting the edge connection them.
            \item If one corresponding vertex is in $T_\theta$ and the other in $C_\theta$, just remove the one vertex in $C_\theta$.
        \end{enumerate}
        Keeping other decisions the same, let the modified decision set be $D_\theta'$.
        \item with probability $1-q_4$, randomly sample any module, say $M_k$, and randomly select any bipartition of its set of core data nodes in $G$, say $X_k^* = X_i^* \cup X_j^*$ and split $M_k$ into two modules, $M_i$ and $M_j$ (note that we have yet to assign an index to either module yet), which, as subsets of $G$, contain nodes that are implied by Definition \ref{moduledef}. Also by Definition \ref{moduledef} and \ref{modulegraphdef}, create an edge in $G_\mathcal{M}$ whenever any module intersects $M_i$ or $M_j$ in $G$, including between $M_i$ and $M_j$ if they are not disjoint. Assign directions to all these undirected edges in a way that matches the directions of edges in the original $G_\mathcal{M}$ except between $M_i$ and $M_j$. This means that, for example, if there is an edge from $M_l$ to $M_k$ in the original $G_\mathcal{M}$, then any undirected edge between $M_l$ and $M_i$ or $M_j$ will inherit the same direction. When this is done, randomly select a pair of indices for $M_i$ and $M_j$ (which implies the direction of the edge between them if present) out of all pairs of indices such that the resulting order can be \textit{adjusted} into a topological ordering of $G_\mathcal{M}'$, reject and sample again if there are no such indices. Call this ordering $S'$, modify the decision for any $\theta$ that belongs to both $M_i$ and $M_j$ according to the following:
        \begin{enumerate}
            \item If $M_k$ is in $T_\theta$, then
            \begin{enumerate}
                \item with probability $q_5$, we let the corresponding vertices of $M_i$ and $M_j$ belong to $T_\theta'$, with each neighbor of $M_k$ in $D_\theta$ randomly divided among them, namely, some are wired to $M_i$ and others to $M_j$.
                \item with probability $1-q_5$, let $M_i$ belongs to $T_\theta'$, and $M_j$ belongs to $C_\theta'$ and its only neighbor randomly chosen.
            \end{enumerate}
            \item If $M_k$ is in $C_\theta$, then let both $M_i$ and $M_j$ be in $C_\theta'$, and let (a randomly chosen) one of them have the same neighbor as $M_k$ in $D_\theta$ and the other one have a randomly selected neighbor in $T_\theta$ with a smaller index. 
        \end{enumerate}
        Keeping other decisions the same, let the modified decision set be $D_\theta'$.
    \end{enumerate}
\end{enumerate} 
\end{defn}

\noindent Note that all the probabilities here, especially $p_i$'s, are subject to changes. Although for some, like $q_2$, a particular value (in the case of $q_2$, $2/3$) might make more sense since the value makes the cardinality of $T_\theta$ and $C_\theta$ martingales with respect to discrete time steps so that we are unlikely to end up with everything in one or the other after a large number of steps. \par

The main challenge in defining this walk is that, when perturbing $G_\mathcal{M}$, we need to make sure the new decision set, $\mathcal{D}'$, gives rise to posteriors as desired under the new directed module graph, $G_\mathcal{M}'$. In Definition \ref{randomwalk} we merely give how $\mathcal{D}'$ and $G_\mathcal{M}'$ are defined without checking if they behave nicely in this way, so there are the following results that address this concern.

\begin{lem} \label{decisionmutuallyconnected}
For any $\theta \in \underline{\Theta}$, the subgraph induced in $G_\mathcal{M}$ by the set of vertices that correspond to vertices in $D_\theta$ is complete.
\end{lem}
\noindent Albeit immediate from Definition \ref{modulegraphdef}, this lemma captures the key observation that enables the following proposition:

\begin{prop} \label{decisionwelldefined}
In both part 2(a) and 2(b) of Definition \ref{randomwalk}, taking $G_\mathcal{M}'$, $S'$, and $\mathcal{D}'$ together does give a posterior $p'$ in the posterior space.
\end{prop}
Overall, we have that
\begin{prop} \label{irreducibility}
Each step in the walks defined is reversible. Thus, the random walk is irreducible.
\end{prop}
\section{Justification of Propositions}
\subsection*{Proposition \ref{any_ordering_works}}
Fix any posterior $p\in \mathcal{P}$ and any ordering $S$, by the definition of $\mathcal{P}$, there exists ordering $S'$ and decision set $\mathcal{D}'$ such that $f_{S'}(\mathcal{D}') = p$ . If $S = S'$ then we are done. If not, since both are topological ordering of $G_\mathcal{M}$, which is acyclic, $S$ and $S'$ must differ in such a way that we may WLOG, through a finite sequence of switching the places of two modules that are not ancestors of one another, obtain $S$ from $S'$. But switching the places of two modules that are not ancestors of one another only alters Algorithm 4 by $permuting$ the update terms in step 3 without $changing$ them (as by an earlier lemma, any two such modules must be disjoint). Thus, we can define $\mathcal{D}$ such that the underlying modules it refers to match those of $\mathcal{D}'$ in ordering $S$ (instead of $S'$) and have $f_{S}(\mathcal{D}) = p$.

\subsection*{Proposition \ref{post_unique}}
Given $G_\mathcal{M}$ and $S$, assume that $\mathcal{D} \neq \mathcal{D'}$ are two decision sets. By assumption, $\mathcal{D}$ and $\mathcal{D'}$ only differ in decisions regarding modules---say $M_i$---where there is some $\theta$ that is in $M_i$ only and no other modules, which implies that we must update $\theta$ in the update term for $M_i$ ($\theta \in U_{M_i}$). This means that the term for $M_i$ obtained using $\mathcal{D}$ and $\mathcal{D'}$ will be different and both will be present in the resulting posterior, hence they are different.

\subsection*{Proposition \ref{decisionwelldefined}}
For 2(a), it is clear that the new decision of any $\theta$ belongs to neither of $M_i$ and $M_j$. For $\theta$ in only one of the modules being merged, WLOG says $M_i$ the place where problems may arise is that we assign a new index $k$ to $M_i$ and this could alter the order of that module in $D_\theta$. But upon a closer look that cannot be: Lemma \ref{decisionmutuallyconnected} states that vertices of $D_\theta$ are mutually connected, so if their order is changed in $D_\theta'$, the assignment of the new index $M_k$ would be impermissible in the first place. For $\theta$ in both $M_i$ and $M_j$, the same lemma implies that successfully assigning the index $k$ and obtaining a topological ordering $S'$ entails $M_i$ and $M_j$ must have adjacent indices in $D_\theta$. This means that the modifying we do to those $D_\theta$ is well-defined.
For 2(b), a similar justification can be executed by symmetry.

\subsection*{Proposition \ref{irreducibility}}
That the walk is reversible is so by design. In Definition \ref{randomwalk}, for each possible perturbation that can happen in a particular step, there is a possible perturbation that directly counteracts it. \par
To prove the walk is irreducible, we need to prove that
\begin{enumerate}
    \item we can always go back to a particular state 
    \item We can always go from that state to any other state
\end{enumerate}
We shall choose the full Bayesian posterior as the referencing state, which, as remarked earlier, is the outcome of having only one module in $\mathcal{M}$. Call it the \textit{null state}. \par
Going from any state to the null state is simple: we can continuously merge modules until there is only one left. The module formation rules in definition \ref{moduledef} ensure that the only module left correctly includes all nodes in $V(G)\setminus \Theta_S$ as desired. \par

Now, the other direction follows from the fact that the walk is reversible: if we can reach the null state from any state $a$ in, say, $n$ steps, then by reversibility we can reach $a$ from the null state in $n$ steps, too.

\end{document}